\documentstyle[aps,floats,epsf,epsfig]{revtex}

\begin{document}
\wideabs{
\title{Quantum Phase Transitions in the Interacting Boson Model: Integrability,
level repulsion and level crossing.}
\author{J.M. Arias$^1$, J. Dukelsky$^2$, J.E. Garc\'{\i}a--Ramos$^3$}
\address {$^1$ Departamento de F\'{\i}sica At\'omica, Molecular y Nuclear,
Facultad de F\'{\i}sica, \\ Universidad de Sevilla, Apartado~1065,
41080 Sevilla, Spain.\\
$^2$ Instituto de Estructura de la Materia, CSIC, Serrano 123,
28006 Madrid,
Spain. \\
$^3$ Departamento de F\'{\i}sica Aplicada, Universidad de Huelva,
21071 Huelva, Spain.}

\maketitle
\begin{abstract}
{We study the quantum phase transition mechanisms that arise in
the Interacting Boson Model. We show that the second-order nature
of the phase transition from U(5) to O(6) may be attributed to
quantum integrability, whereas all the first-order phase
transitions of the model are due to level repulsion with one
singular point of level crossing. We propose a model Hamiltonian
with a true first-order phase transition for finite systems due to
level crossings.}

\noindent PACS number: 21.60.Fw, 21.10.Re, 64.60.Fr

\end{abstract}}

Quantum phase transitions (QPTs) have attracted great attention
from the theoretical and experimental communities in recent years.
Experiments on high-$T_{c}$ superconductors, on quantum-Hall
systems, on Bose or Fermi dilute gases in different trap
geometries and on nuclei close to a critical point challenge
theoreticians to develop reliable approaches with which to
describe the critical properties of these systems.

A QPT describes a structural change in the properties of the
ground state of the system associated with the variation of a
control parameter (the temperature being the usual one in
classical phase transitions). This parameter might be the level of
doping in high-$T_{c}$ superconductors, the magnetic field in
quantum-Hall systems, the scattering length in dilute gases, the
nucleon number in nuclear phase transitions, etc...

The phase diagram of the Interacting Boson Model (IBM)\cite{Ia} of
nuclei has been studied at the mean-field level\cite{Die,Feng},
using both Catastrophe theory\cite{Cas} and the Landau theory of
phase transitions\cite{Jo1}. Both treatments show that the model
displays first-order phase transitions from spherical to deformed
shapes and from oblate to prolate deformed shapes, with an
exceptional point of an isolated second-order phase transition. In
this work, we study in detail the phase transitions associated
with the IBM.

The IBM is a phenomenological model of nuclear structure with a deep connection to the underlying
microscopic Shell Model. In this model, which has been very successful in describing nuclear
spectroscopic properties of collective nuclei, pairs of correlated nucleons of angular momentum
$L=0$ and $L=2$ are represented by $s$ and $d$ bosons, respectively. In its simplest version the
model does not distinguish between protons and neutrons and has an underlying group structure based
on the dynamical group U(6).

The most general IBM Hamiltonian can be written in terms of $6$ free parameters, as a linear
combination of the linear and quadratic Casimir operators of $U(5)$ and the quadratic Casimir
operators of $O(6)$, $SU(3)$, $O(5)$ and $O(3)$. A convenient and frequently-used form of the IBM
Hamiltonian that keeps all the main ingredients related to the structure of the ground state is the
Consistent-Q Hamiltonian \cite {Ca}

\begin{equation}
H= x n_{d}+\frac{x -1}{N}Q^{\chi }\cdot Q^{\chi } ~,  \label{HQ}
\end{equation}
where $n_{d}=\sum_{\mu }d_{\mu }^{\dagger }d_{\mu }$, $Q^\chi_{\mu
}=\left[ d^{\dagger }\widetilde s+s^{\dagger }\widetilde d\right]
_{\mu }^{2} +\chi \left[ d^{\dagger }\widetilde{d}\right] _{\mu
}^{2}$ and $N$ is the total number of bosons, which is equal to
the number of nucleon pairs in the valence space.

One of the most important features of the IBM is the existence of
four distinct dynamical symmetries (DS), each representing a
well-defined phase of nuclear collective motion. A quantum system
has a DS if the Hamiltonian can be expressed as a linear
combination of the Casimir operators of a subgroup chain of the
dynamical group. The four dynamical symmetries are: the $U(5)$
symmetry for spherical vibrational nuclei ($x =1$), the $SU(3)$
symmetry for prolate deformed nuclei ($x =0$, $\chi=-\sqrt{7}/2$),
the $O(6)$ symmetry for $\gamma $ unstable deformed nuclei ($x
=0$, $\chi =0$) and the $\overline{SU(3)}$ symmetry for oblate
deformed nuclei ($x =0$, $\chi =\sqrt{7}/2$).

\begin{figure}
\hspace{1cm} \epsfysize=6cm
\epsffile{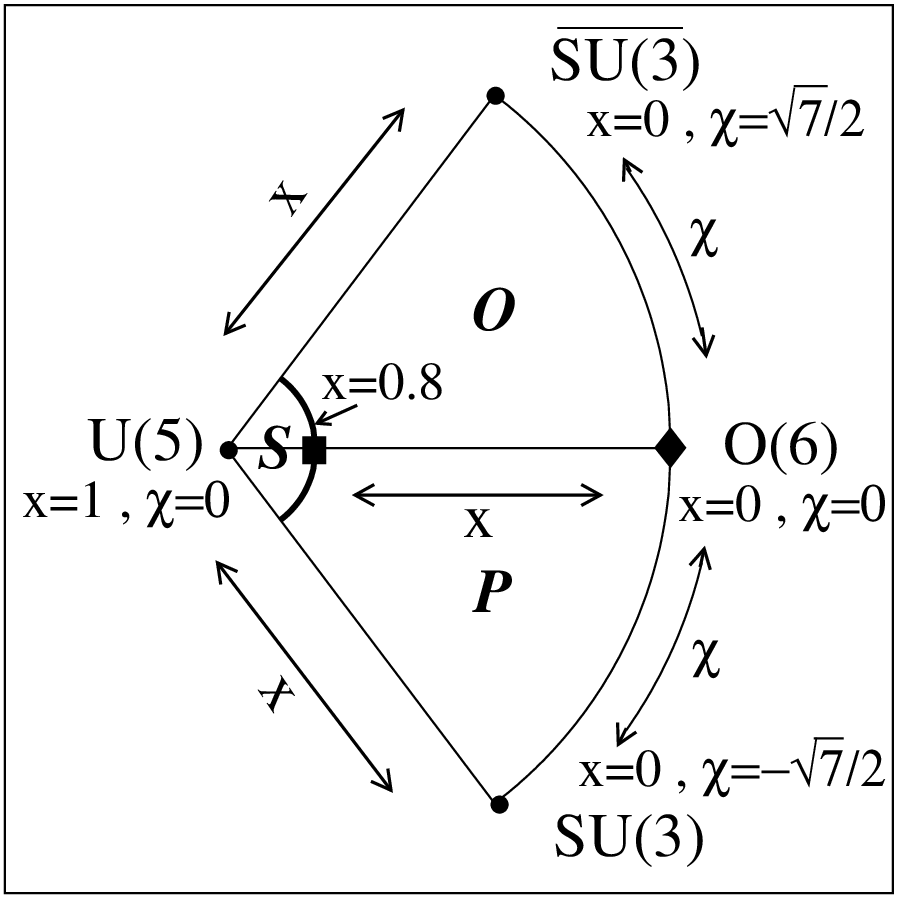} \vspace{0.5cm}
 \caption{Phase diagram of the Interacting Boson Model in the
phase space of the control parameters $x$ and $\chi$. S,
  O, and P stand for the spherical, oblate, and prolate phases respectively.}
\label{fig1}
\end{figure}

The symmetries of the IBM and the transitions between them are
illustrated in the extended Casten triangle \cite{Jo2} of Figure
\ref{fig1}, along with the corresponding values of the parameters
$x$ and $\chi $. Distinct phases associated with three ``shapes",
denoted S for spherical, P for prolate deformed and O for oblate
deformed, can be seen in the figure.  The three different shape
phases are separated by two lines of first-order phase
transitions,  $x=0.8$ and $\chi=0$, respectively, with the
exception of a triple point\cite{Jo1} at $x=0.8$ and $\chi=0$,
which is represented by a solid square and for which the phase
transition is second order. Recently, two effective models based
on the Bohr Hamiltonian, called  $X(5)$ and $E(5),$
have been proposed to describe the physics at the critical points along the $%
U(5)-SU(3)$ and $U(5)-O(6)$ lines, respectively \cite{Ia2}.

The aim of this letter is to go beyond the mean-field description
of the phase diagram in Fig.~\ref{fig1}. We will show that the
first-order QPTs arise due to level repulsion between two states
with different properties, as was earlier recognized in
\cite{Jo3}. However, there are two exceptional singular critical
points, represented by the bold square and the bold diamond in
Fig. 1, for which the mechanism triggering the phase transition is
completely different.

We begin with a discussion of the transition from $O(6)$ to $%
U(5)$. It is described within the Hamiltonian (\ref{HQ}) by
setting $\chi =0$ and varying $x $ from $0$ to $1$. Earlier
investigations of the properties of the level spacings along this
leg of the Casten triangle showed a Poisson distribution. This
indicates that the model is integrable along this leg \cite{Alha},
even though it is not characterized by a global Dynamical
Symmetry.

Quantum integrability requires the existence of a complete set of
mutually commuting hermitian operators. These operators are the
integrals of motion or quantum invariants and their eigenvalues
are the conserved quantities that completely label a unique basis
of common eigenstates. A DS is a particular class of integrable
model in which each one of the Casimir operators in the subgroup
decomposition chain is a quantum invariant. But there are also
other integrable and exactly solvable models that have played
important roles in the understanding of quantum many-body systems.
Examples include the Heisenberg and Hubbard models, both of which
are solvable by the Bethe ansatz. The Hamiltonian (\ref{HQ}) for
$\chi=0$ is a particular case of a general class of pairing models
that have been shown to be quantum integrable \cite{duk1}.  The
quantum invariants for boson pairing Hamiltonians and the
corresponding exact solutions were discussed in Ref. \cite{duk2}.
Here we will specify the form of the quantum invariants in the
$O(6)-U(5)$ transition. Since we are only interested in properties
of the ground state, we focus our treatment on the case in which
all bosons are paired to zero angular momentum.

\begin{figure}
\epsfysize=6cm \epsfxsize=8.5cm \epsffile{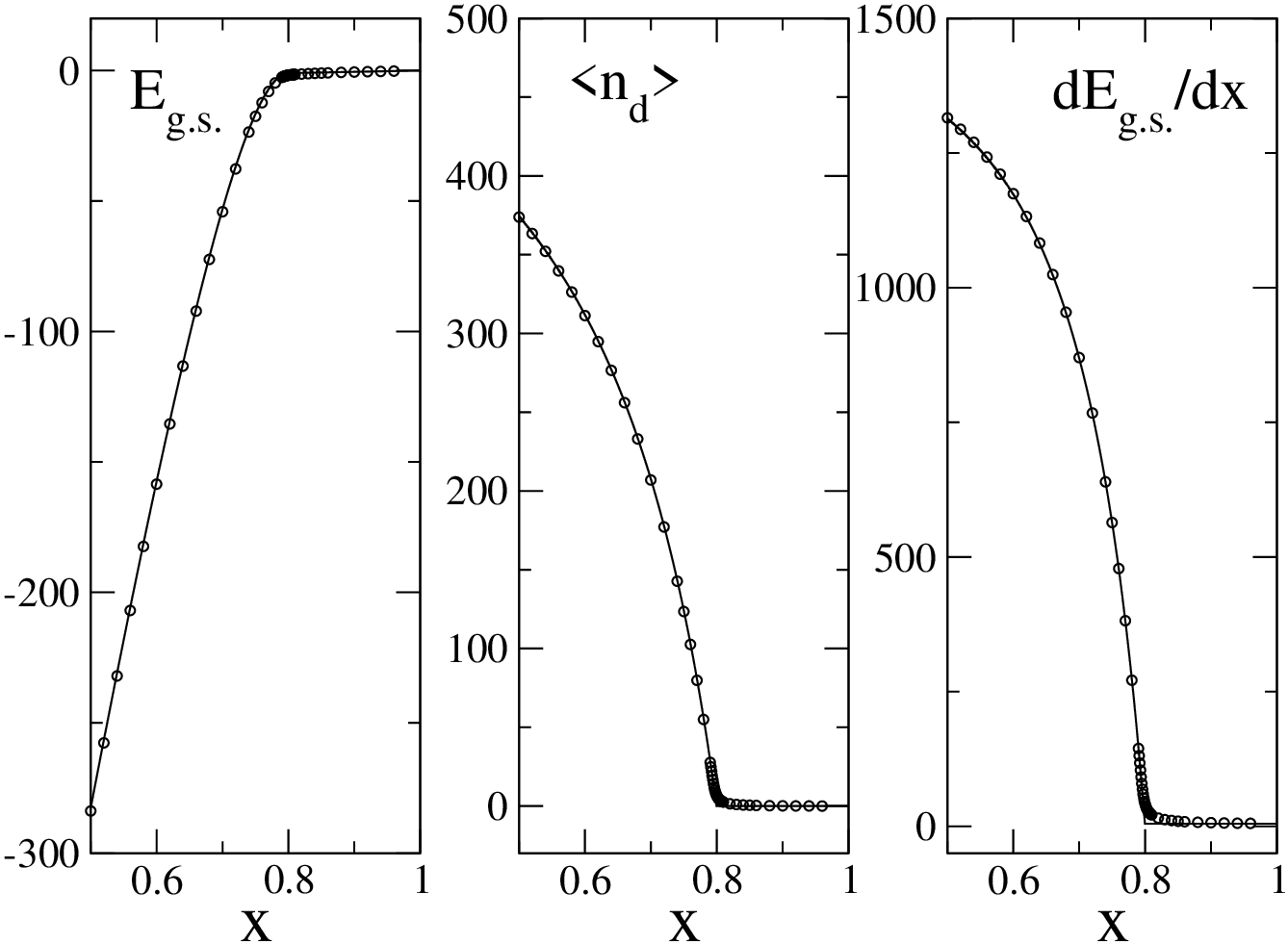}
\vspace{0.5cm}
 \caption{Transition from $O(6)$ to $U(5)$. The ground
  state energy (left panel), expectation value of the $n_d$
 (central panel), and derivative of the ground state energy (right
  panel) are presented as a function of the control parameter
  $x$ for a system with 1000 bosons. The exact results are represented by circles
  and the mean field results by a full line.}
\label{fig2}
\end{figure}

We first note that this transition is embedded in the decomposition $%
U(6)\supset \cdots \supset O(5)\supset O(3)\supset O(2)$, where a
missing parameter dependent ``symmetry'' or quantum invariant
commuting with the Casimir operators of all the groups in the
chain is responsible for the integrability of the model.

As noted in references \cite{Alha,cej}, the Hamiltonian itself, interpolating between the Casimir
operators of $U(5)$ and $O(6)$, constitutes the additional independent quantum invariant ensuring
the integrability of the system.

The eigenvalues of the Hamiltonian (\ref{HQ}) for $\chi=0$ are \cite{duk2}

\begin{equation}
E= \frac{20 (x -1)}{N} \sum_{\alpha =1}^{N/2}\frac{1}{%
2-e_{\alpha }} ~, \label{rd}
\end{equation}
where the pair energies $e_{\alpha }$ are the solutions of the coupled set of nonlinear Richardson
equations

\begin{equation}
\frac{x N}{2(1-x)} -\frac{1}{e_{\alpha }}+\frac{5}{2-e_{\alpha }}-4\sum_{\beta \left(
    \neq \alpha \right)
}\frac{1}{e_{\alpha }-e_{\beta }}=0 ~. \label{Rich}
\end{equation}

The Hamiltonian eigenvalues (\ref{rd}) are analytic functions of the control parameter $x$.
Therefore, in the absence of level crossings at the ground state level, a first-order phase
transition is precluded. Level crossings can occur only for integrable models with more than one
parameter dependent integral of motion, an example being the $sdg$ boson model\cite{duk3}.

Because of the exact solvability of the model, we are able to
obtain exact results numerically for a large number of bosons. In
Fig.~\ref{fig2}, we show results for 1000 bosons and compare them
with the mean-field solution. In the left panel, we plot the
ground-state energy versus the control parameter $x$; in the
central panel, the expectation value of the number of $d$ bosons
$<n_d>$ is presented as a function of $x$; in the right panel, we
plot the derivative of the energy versus $x$. Due to the exact
solvability of the model, we can make use of the Hellmann-Feynman
theorem to express the first derivative of the energy as the
expectation value of the derivative of the Hamiltonian (\ref{HQ})
with respect to $x$, {\it i.e.} $\frac{\partial H}{\partial x} =
n_d + \frac{1}{N} Q^0 \cdot Q^0$. Both operators, $n_d$ and
$\frac{\partial H}{\partial x}$, are zero in one phase
(disordered) and different from zero in the other phase (ordered),
fulfilling the requirements for being order parameters. Both
expectation values are continuous at the transition point
satisfying both the Landau and Ehrenfest definitions of a
second-order phase transition.

\begin{figure}
\epsfysize=5cm \epsfxsize=8.5cm \epsffile{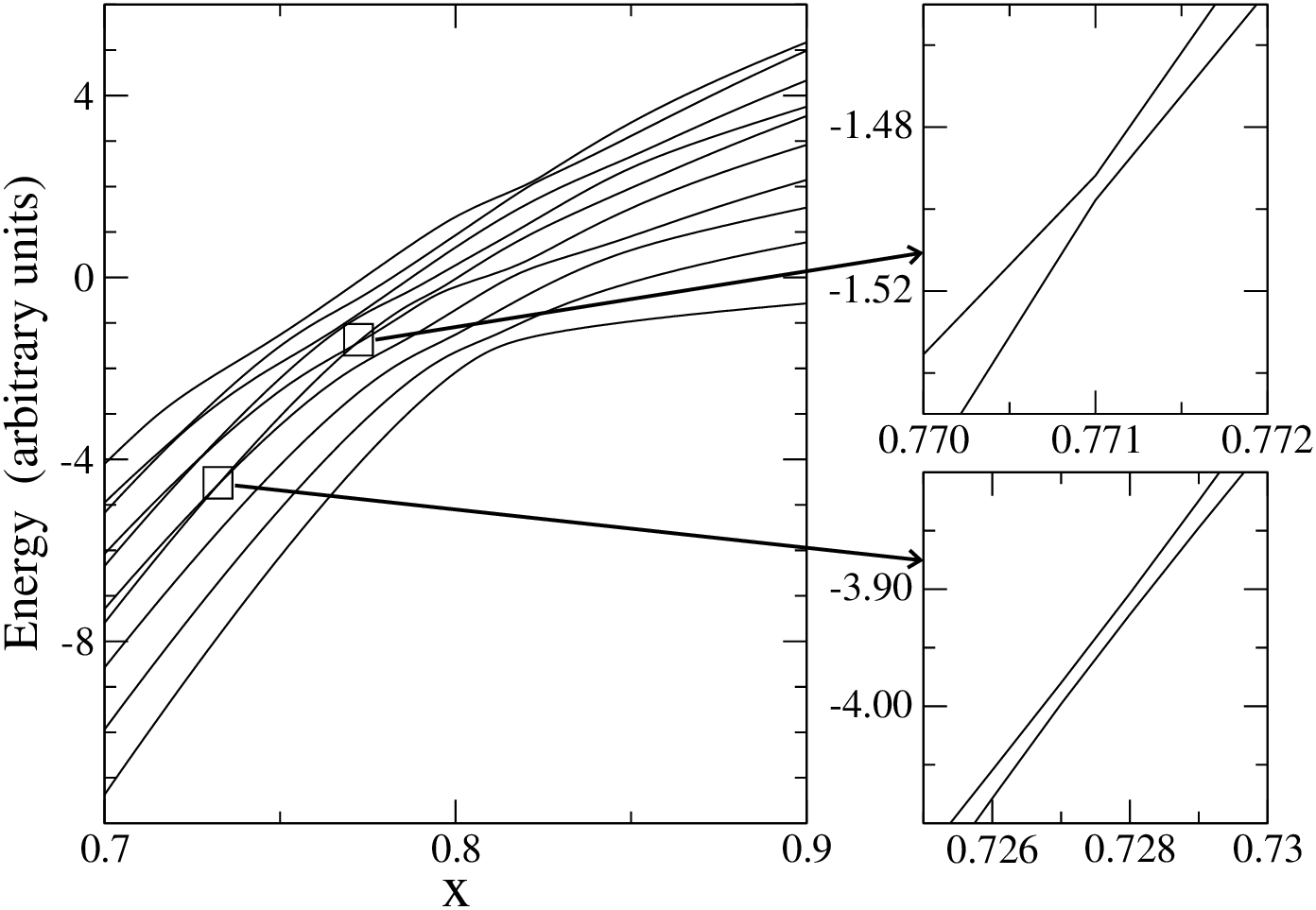}
\vspace{0,5cm}
 \caption{Ten lowest $0^+$ states
in the transition from $U(5)$ to $SU(3)$ for $N=50$ bosons as a
function of $x$. On the right panels we show in an enlarged scale
two examples of level repulsion.} \label{fig3}
\end{figure}

We will now discuss the phenomenon of level repulsion for non-integrable systems. For an arbitrary
finite value of $\chi$, we select a value of $x=x_0 $ in the vicinity of which we can use
non-degenerate perturbation theory. Defining $x'=x-x_0$, we can rewrite the Hamiltonian as $H=x_0
n_{d}+\frac{x_0-1}{N}Q^{\chi}\cdot Q^{\chi}+x'\left( n_{d}+\frac{1}{N}Q^{\chi }\cdot Q^{\chi
}\right) =H_{0}+x' V ~.$ The first and second derivatives of the Hamiltonian eigenvalues in
perturbation theory are\cite{alt} :

\begin{equation}
\frac{\partial E_{n}}{\partial x' }=\frac{\partial E_{n}}{\partial x }=V_{nn}\left( x_0 \right)
~,\label{DE}
\end{equation}

\begin{equation}
\frac{\partial ^{2}E_{n}}{\partial x ^{'2}}=
\frac{\partial ^{2}E_{n}}{\partial x ^{2}}=2\sum_{m\neq n}\frac{%
V_{nm}^{2}\left( x_0 \right) }{E_{n}\left( x_0 \right) -E_{m}\left( x_0 \right) } ~. \label{D2D}
\end{equation}

Assuming that there is a pair of energy levels that get very close
in energy, the energy gap between them is $\Delta \left( x \right)
=E_{2}\left( x \right) -E_{1}\left( x \right)$. Neglecting the
contribution of all other states, perturbation theory for the gap
derivative gives

\begin{equation}
\frac{\partial \Delta }{\partial x }=V_{22}-V_{11} , \; \; \; \; \frac{\partial ^{2}\Delta
}{\partial x ^{2}}=\frac{4V_{12}^{2}}{\Delta } ~. \label{derd}
\end{equation}

We can interpret $E_{1}$ and $E_{2}$ as the positions of two
fictitious classical particles in a one-dimensional space. Then
(\ref{derd}) are the equations for their relative velocity and the
force between them. Clearly, if the interaction matrix element
$V_{12}$ remains finite as $\Delta \rightarrow 0$, there is an
infinite repulsion and this prevents the particles from colliding.
This is a pictorial view of the phenomenon of level repulsion for
non-integrable Hamiltonians. On the other hand, if there is an
additional conserved quantity and both levels have different
symmetries, then $V_{12}$ is strictly zero and the two levels can
cross. In what follows, we discuss the application of these ideas
first to the $U(5)$ to $SU(3)$ transition and then to the
$\overline{SU(3)}$ to $SU(3)$ transition.

In Fig. \ref{fig3}, we show in the left panel the ten lowest
$0^{+}$ states for a system of $50$ bosons as a function of $x$
along the $U(5)$ to $SU(3)$ transition. There are several cases in
which two $0^{+}$ levels get very close in energy but, as just
discussed, the closer they get the larger the level repulsion
between them is. In the right panels, we magnify two of the cases
of closest approach  to show that indeed the levels repel and do
not cross. In particular, there is a smooth crossover in the
ground-state energy at the point of closest approach with the
first excited state. The dynamics of these two levels is well
described by Eq. (\ref{derd}) with finite matrix elements
$V_{ij}$. The crossover will change gradually as the number of
bosons increases towards a non-analytic point in the thermodynamic
limit, defining a first-order phase transition. From the above, we
conclude that level repulsion is indeed responsible for the
first-order phase transition in this case\cite{Jo3}.

Now we turn to the third leg of the triangle ($x=0$), the
transition from deformed oblate ($\overline{SU(3)}$ shapes,
$\chi=\sqrt{7}/2$) to deformed prolate ($SU(3)$ shapes,
$\chi=-\sqrt{7}/2$) shapes, passing through the $\gamma$-unstable
$O(6)$ case with $\chi=0$. In this case, the first-order phase
transition takes place precisely at the $O(6)$ dynamical
symmetry\cite{Jo2}. At this critical point, the
$0^{+}$ states can be classified with the $\tau $ quantum number of an $%
O(5) $ symmetry. As discussed above, the realization of this new symmetry allows the crossing of
levels with different quantum numbers. The Hamiltonian (\ref{HQ}) at this point is

\begin{equation}
H=-\frac{1}{N}Q^{0}\cdot Q^{0}=\frac{1}{2N}\left( C_{2}\left[ O(6)\right] -C_{2}\left[ O(5)\right]
\right) ~,  \label{Cas}
\end{equation}
with eigenvalues $E_{\sigma \tau }=\left[ -\sigma \left( \sigma +4\right) +\tau \left( \tau
+3\right) \right] /2N$. The lowest $O(6)$ band corresponds
to $\sigma =N$ and $\tau =0,3,\cdots $. In the large $N$ limit, the $%
O(5)$ states within the ground state $O(6)$ band will be
degenerate, since the energy spacings go as $N^{-1}$. According to
the above analysis, in the thermodynamic limit an infinite set of $%
0^{+}$ states with different seniority quantum numbers $\tau $
cross. The crossing is allowed because of the $O(5)$ symmetry at
the critical point, implying that the interaction matrix elements
in (\ref{derd}) are zero.

\begin{figure}
\epsfysize=5cm \epsfxsize=8.5cm \epsffile{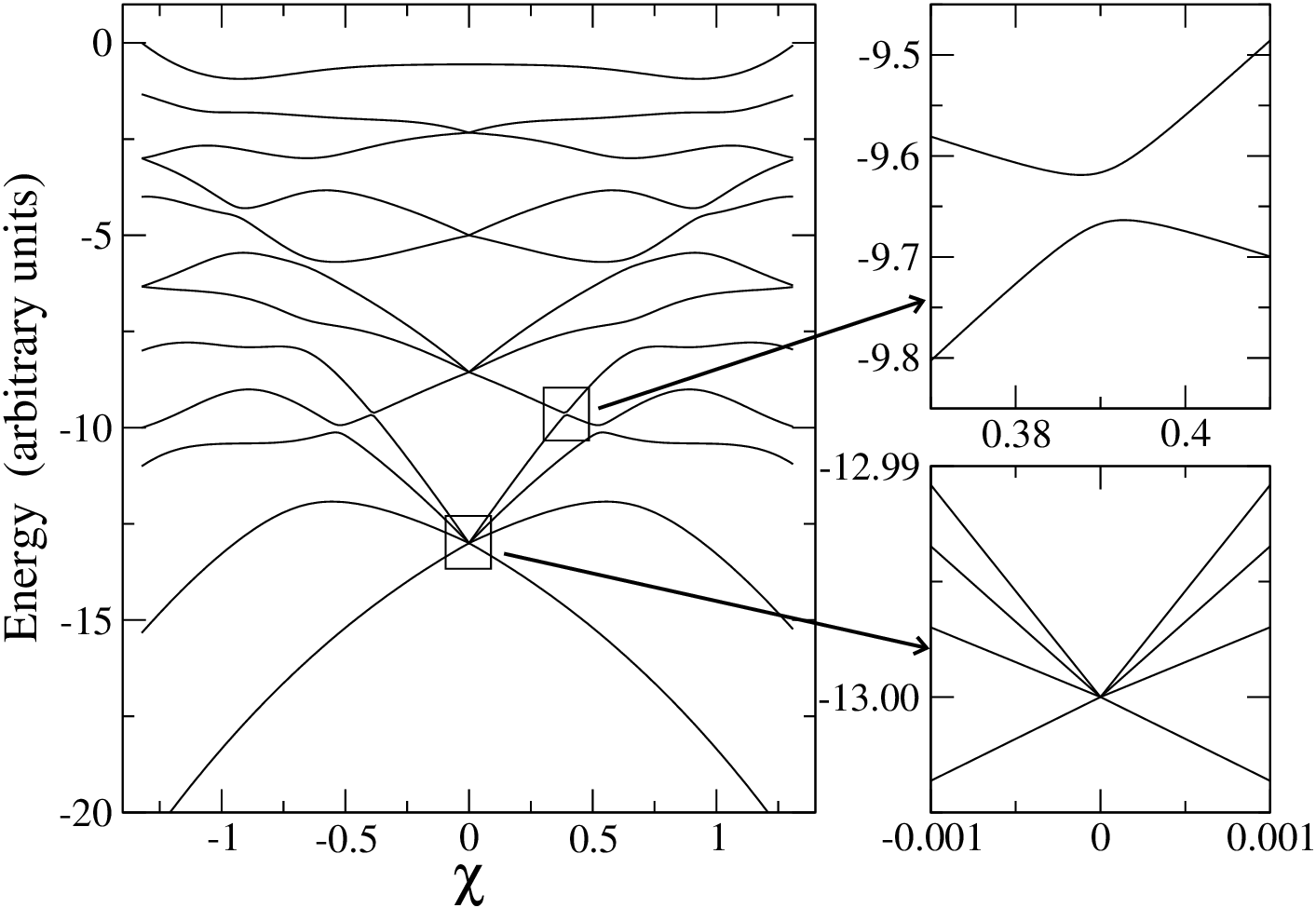}
\vspace{0,7cm} \caption{Transition from $\overline{SU(3)}$ to
$SU(3)$ through the
  $O(6)$ limit. All $0^+$ eigenstates of the Hamiltonian (\ref{Hret}) for
  a system with $N=9$ bosons are represented as a function of the
  control parameter $\chi $. The right panels show in a larger
  scale the level repulsion of two close levels at $\chi \neq 0$ (right-top panel) and the level
  crossing at $\chi=0$
  (right-bottom panel).}
\label{fig4}
\end{figure}

As just noted, the exact degeneracy in the levels of the ground-state band at $x=0$ and $\chi=0$
only occurs in the large-N or thermodynamic limit. To better visualize this QPT in the context of
finite systems, we therefore consider a hamiltonian similar to that in  (\ref{HQ}), but with an
additional term that cancels the $O(5)$ term in it that produces energy spacings within the
ground-state band for finite N (see eq. (\ref{Cas})).  With this in mind, we consider the following
Hamiltonian associated with the $x=0$ transition leg:

\begin{equation}
H=-\frac{1}{N}\left[ Q^{\chi }\cdot Q^{\chi }+\left( \frac{1}{2}-\frac{%
\left| \chi \right| }{\sqrt{7}}\right) C_{2}\left[ O(5)\right] \right]. \label{Hret}
\end{equation}

For $\chi =$ $-\frac{\sqrt{7}}{2}$, the Hamiltonian (\ref{Hret})
is in the $SU(3)$ limit;
for $\chi =$ $\frac{\sqrt{7}}{2}$, it is in the $\overline{SU(3)}$ limit; and for $%
\chi =0$ it is proportional to the quadratic Casimir operator of
the $O(6)$ group. The second term in the modified Hamiltonian
(\ref{Hret}) thus cancels out the $O(5)$ breaking of the
degeneracies in the ground-state band that occurs for finite N.

In Fig.~\ref{fig4}, we show all of the $0^{+}$ eigenstates of the
Hamiltonian (\ref{Hret}) for a system with $N=9$ bosons as a
function of $\chi $. At $\chi =0$, the four levels that correspond
to the ground-state band with $\sigma=9$ cross. This crossing
gives rise to a first-order QPT for a ``finite'' system. This is
the unique point along the $x=0$ leg at which level crossings are
allowed due to the realization of the $O(5)$ symmetry. Away from
this point the mechanism of level repulsion sets in. In the upper
right panel of Fig. \ref{fig4}, we show in an enlarged scale an
apparent level crossing at $\chi\ne 0$. The mechanism of level
repulsion predicted in eq. (\ref{derd}) is numerically verified.
On the contrary at $\chi = 0$, the realization of the $O(5)$
symmetry allows the crossing of levels with different $\tau$
quantum numbers, as shown in the lower right panel.

In summary, there are two singular points of QPTs in the IBM phase
diagram. The first is in the transition from $U(5)$ to $O(6)$ and
corresponds to a second-order QPT due to the integrable nature of
the Hamiltonian along this line, which ensures that the
ground-state energy is an analytic function of the control
parameter. The second singular point is located at the QPT from
$SU(3)$ to $\overline{SU(3)}$, precisely at the $O(6)$ critical
symmetry. In the thermodynamic limit, this QPT is associated with
the crossing of an infinite number of $0^+$ states. By adding to
the Hamiltonian (\ref{Cas}) a term proportional to the $O(5)$
Casimir operator, we were able to realize this QPT for a finite
system. This is an unusual, but especially interesting, case to
study further, since most phase transitions take place in the
thermodynamic limit. Therefore, the model described by the
Hamiltonian (\ref{Hret}) may be useful to explore the universal
properties of first-order phase transitions.

These two isolated critical points represent exceptional points in
the phase diagram of the IBM. Along the lines separating the
spherical, prolate and oblate phases, the mechanism underlying the
first-order phase transitions is level repulsion between the
ground state and the first excited state. For finite $N$, this
mechanism is reflected by a soft crossover, but in the large $N$
limit the model develops a singularity in the derivative of the
gap consistent with a first-order phase transition. This
conclusion is also valid for more general IBM hamiltonians, since
level repulsion is a universal phenomenon associated with quantum
non-integrability.

%
This work was supported in part by the Spanish DGI under projects
numbers BFM2002-03315 and BFM2000-1320-C02-02.  We acknowledge
useful discussions with S. Pittel, P. Van Isacker, J. Jolie and F.
Iachello.


\end{document}